# Magnetism and Magnetic Anisotropy of Transition Metal-Phthalocyanine Molecules

Jun Hu and Ruqian Wu

*Department of Physics and Astronomy, University of California, Irvine, California 92697-4575, USA*

Search for single-molecule magnets (SMMs) with high blocking temperature ($T_B$) is urgent for practical applications in magnetic recording, molecular spintronics and quantum computing. Based on the First-principles calculations, magnetic anisotropy energies (MAE) of the transition metal-Phthalocyanine (TM-Pc) molecules are investigated and the mechanism that determines the MAE of TM-Pc molecules is established. In particular, colossal MAE > 100 meV can be obtained by adding an Os atom on RuPc and OsPc, so these molecules may offer ultrahigh thermal stability in devices.

Extensive studies of single-molecule magnets (SMMs) have evoked the emergence of a new research field: molecular spintronics, where SMMs are used as the core building blocks for recording, transport and sensing devices [1,2,3]. Typical SMMs comprise of organic molecules and 3d transition metal (TM) cores (especially Mn, Fe and Co) and their electronic and magnetic properties can be conveniently tuned by selecting appropriate molecules or TM atoms [1, 2, 3, 4, 5]. Among all SMMs, the TM-Phthalocyanine (notated as TM-Pc) molecules have been mostly studied, with a research history back to 100 years ago [6,7]. Aside from their promising electronic and magnetic properties, TM-Pc molecules have simple atomic structures as depicted in Fig. 1(a) and high chemical and thermal stability in ambient [8]. In addition, TM-Pc molecules may show interesting features when they are in contact with substrates. For instance, a MnPc molecule on Pd(111) have two different magnetic ground states because of the competition between Kondo screening and superconducting pair-breaking interactions [9]. MnPc and CoPc molecules couple to a magnetic substrate through RKKY interaction [10] and form an antiferromagnetic one-dimensional chain on Pb(111) [11]. FePc molecules were found to switch their magnetic anisotropy to the perpendicular direction on an oxidized Cu(110) surface [12,13]. Furthermore, the electronic and magnetic properties of TM-Pc molecules can be modified upon adsorbing adatom [14] or small molecule such as CO [15].

The main obstacle for the development of TM-Pc molecular devices is their low blocking temperature, $T_B$, a quantity that denotes the threshold temperature for holding their spin orientations against thermal fluctuations [1]. Fundamentally, $T_B$ scales with the magnetic anisotropy energy (MAE), and typical $T_B$ of 3d TM-Pc molecules is less than 10 K (or equivalently, MAE ~ 1 meV) [16,17,18,19]. To keep stable magnetization of molecules up to room temperature (RT) for practical applications, it is crucial to find SMMs with $E_{MCA}$ larger than 30 or even 50 meV. Obviously, we need to seek potential candidates from 4d or 5d TM-Pc molecules that have large spin-orbit coupling (SOC) strength. Although all 4d and 5d TM-Pc molecules have been synthesized [20,21], studies for their magnetic properties are rather rare.

In this paper, the electronic and magnetic properties of all TM-Pc molecules are studied through density functional theory (DFT) calculations. We found that only WPc and RePc have moderate positive MAE ~ 22 meV. However, placing an additional Os atom on RuPc and OsPc molecules can produce colossal MAEs as large as 223 and 136 meV, respectively. Furthermore, both new molecules (Os/RuPc and Os/OsPc) are structurally stable. Therefore, they should be easy to fabricate and useful for various molecular spintronics applications.

Our DFT calculations were carried out with the Vienna ab-initio simulation package (VASP) [22,23]. The interaction between valence electrons and ionic cores was described within the framework of the projector augmented wave (PAW) method [24,25]. The spin-polarized generalized gradient approximation (GGA) was used for the exchange-correlation potentials and the effect of spin-orbit coupling was invoked self-consistently [26]. The energy cutoff for the plane wave basis expansion was set to 400 eV. Periodic boundary condition was used with a large unit cell that ensures the distance between two neighboring molecules larger than 15 Å, sufficient to mimic the environment for single molecules. The atomic positions were fully relaxed with a criterion that requires the force on each atom smaller than 0.01 eV/Å.

We adopted the Torque method proposed by Wang et al [27,28], for the determination of uniaxial magnetic anisotropy energy,

$$MAE = \sum_{i \in occ} \left\langle \psi_i \left| \frac{\partial H_{SO}}{\partial \theta} \right| \psi_i \right\rangle_{\theta=45°}. \qquad (1)$$

Here, θ is the polar angle away from the molecular axis for spin momentum, $\Psi_i$ is the $i$th relativistic eigenvector, and $H_{SO}$ is the SOC Hamiltonian [29]. Recently, we implemented this method in the framework of pseudo-potential PAW method. To check the reliability of the torque approach for cases with strong SOC, we also used the direct method to calculate the MAE as [30,31]

$$MAE = E(\theta=90^0) - E(\theta=0^0). \qquad (2)$$

Here, $E$ represents the total energy of self-consistent calculations with the inclusion of SOC for each spin orientation. Note that the direct method is much more expensive than the torque method due to the need of large number of k points in the Brillouin zone for periodic systems [27,28,29]. Furthermore, the torque method allows rigid band model analysis for the prediction of MAE against the shift of the Fermi level, and also the decomposition of MAE into contributions from different spin channels, atoms and electronic states. For the convenience of discussions, we use MAE(uu), MAE(dd) and MAE(ud+du) to denote the contributions from SOC interactions between the majority spin states, minority spin states, and cross spin states, respectively.

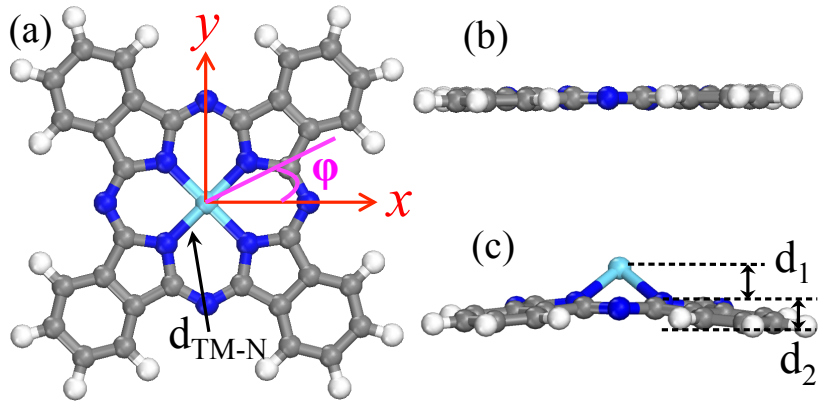

**FIG. 1.** Top and side views of a transition metal phthalocyanine (TM-Pc) molecule. (b) and (c) are for planar and non-planar TM-Pc molecule, respectively. The cyan, blue, grey and white spheres stand for TM, N, C and H atoms, respectively.

After structural relaxation, we found that most TM-Pc molecules prefer a planar geometry as shown in Fig. 1(a) and 1(b), whereas group IIIB and IVB cores (except Ti) tend to stay out of the molecular plane as depicted in Fig. 1(c). In all planar molecules, the TM-N bond lengths ($d_{TM-N}$) are $1.95 \pm 0.05$ Å, regardless the size of core atoms due to the constraint of the macrocycle. For the non-planar molecules, however, $d_{TM-N}$ may vary from 2.09 Å to 2.35 Å. Moreover, their macrocycles also deform as described by non-zero $d_2$, the height of innermost N atoms above the molecular base-plane in Fig. 1(c). The largest $d_1$ and $d_2$ (1.6 Å and 0.5 Å, respectively) were found in LaPc. Nevertheless, the deformed structure for these elements is only the precursor of the more stable bis(Phthalocyaninato)-TM structure (TM-Pc$_2$), as extensively discussed for YPc$_2$ and TbPc$_2$ [18,21].

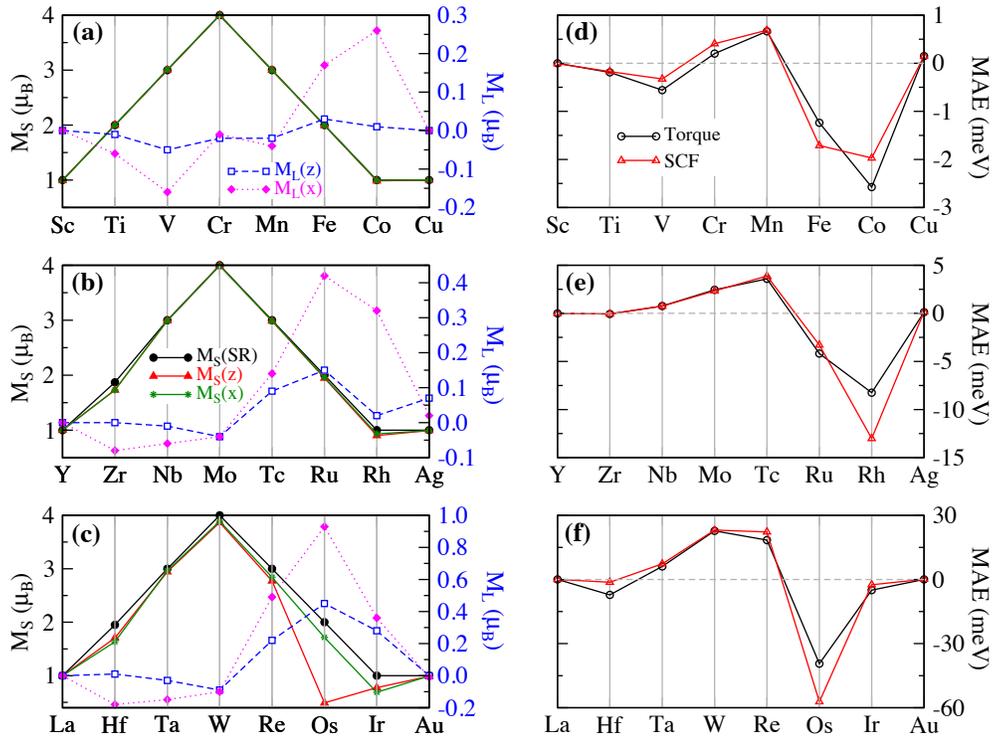

**FIG. 2.** (a)-(c) Spin moment ($M_S$) and orbital moment ($M_L$) from nonrelativistic calculations (SR) and relativistic calculations with spin orientation along $z$ or $x$ axis. (d)-(f) Magnetic anisotropy energy (MAE) of the TM-Pc molecules. Both the torque method and the full self-consistent calculations (SCF) with SOC were used to calculate the MAE.

All core atoms in the TM-Pc molecules adopt the charge state of +2 (denoted as TM$^{2+}$), i.e., the two $s$ electrons are donated to the neighboring N atoms. As a result, their electronic configuration can be denoted as $d^n$. Moreover, the D$_{4h}$ symmetry around the TM cores splits the $d$ orbitals into four groups: $b_{1g}$ (xy) and $b_{2g}$ (x$^2$-y$^2$) for the in-plane components, along with $a_{1g}$ ($z^2$) and $e_g$ (xz and yz) for the out-of-plane components with the coordinate system given in Fig. 1(a). DFT calculations indicate that these orbitals actually intermix with each other and also with N-2p states, and their energy sequences and separations sensitively depend on the core atoms [32]. In general, the energy level of the $b_{1g}$ orbital is higher than that of any other orbital So the spin magnetic moment ($M_S$) of a TM-Pc molecule follows a linear rule: $M_S = n\ \mu_B$ ($n \leq 4$); $M_S = (8 - n)\mu_B$ ($5 \leq n \leq 8$); or $M_S = (10 - n)\mu_B$ (n=9 or 10), as shown by the black lines in Fig. 2(a)-(c). Consequently, the largest $M_S$ is 4.0 $\mu_B$ for molecules in the $d^4$ configuration (CrPc, MoPc and WPc), while NiPc, PdPc, PtPc and IIB-Pc (ZnPc, CdPc and HgPc) molecules are nonmagnetic. When the SOC effect is invoked, the values of $M_S$ of most 4d and 5d TM-Pc molecules decrease, but those of the 3d TM-Pc molecules remain almost unchanged, as shown in Fig. 2(a)-(c). In addition, we found that the values of $M_S$ for in-plane [$M_S(x)$] and perpendicular [$M_S(z)$] magnetizations are almost the same for all cases except OsPc of which the $M_S(z)$ is much smaller than the $M_S(z)$ (0.5 $\mu_B$ v.s. 1.7 $\mu_B$), manifesting the giant spin anisotropy of OsPc.

The calculated MAEs with both the torque and direct methods are plotted in Fig. 2(d)-(f). Interestingly, these two methods produce the same trend of MAEs for all TM-Pc molecules as the atomic number of the core atom changes. In fact, the two approaches produce almost the same MAEs for most cases with only a few exceptions such as FePc, CoPc, RhPc, and OsPc. It appears that the perturbative torque method is very reliable for the determination of magnetic anisotropy of TM-Pc molecules, even the charge and spin densities are frozen during the reorientation of magnetization, and we will use results obtained from the torque method in the following discussions for the easiness of analyses.

Clearly, 3d-, 4d- and 5d-Pc molecules follow the similar trend in the *MAE ~ n* dependence, with an exception VPc. First, all $d^1$ (Sc, Y and La) and $d^9$ (Cu, Ag and Au) TM-Pc molecules have negligible MAEs, which means that they have no obviously preferential direction of magnetization. The $d^2$ (Ti, Zr and Hf) TM-Pc molecules slightly prefer in-plane magnetization, manifested by their small negative MAEs. The MAEs become positive for the $d^3$ (Nb and Ta), $d^4$ (Cr, Mo and W) and $d^5$ (Mn, Tc and Re) TM-Pc molecules, so they have perpendicular magnetization. The MAEs turn to negative again for the $d^6$ (Fe, Ru and Os) and $d^7$ (Co, Rh and Ir) TM-Pc molecules and the largest magnitude is 57 meV for OsPc (from the direct approach). Nevertheless, since the in-plane anisotropy with the change of azimuthal angle is relatively small (e.g., 8 meV for OsPc with an easy axis that is 45° away from the x-axis), only molecules with large positive MAEs are useful to withhold thermal fluctuation. To this end, it appears that only WPc and RePc are suitable candidates among neutral TM-Pc molecules, with MAEs of 23 meV and 22 meV, respectively.

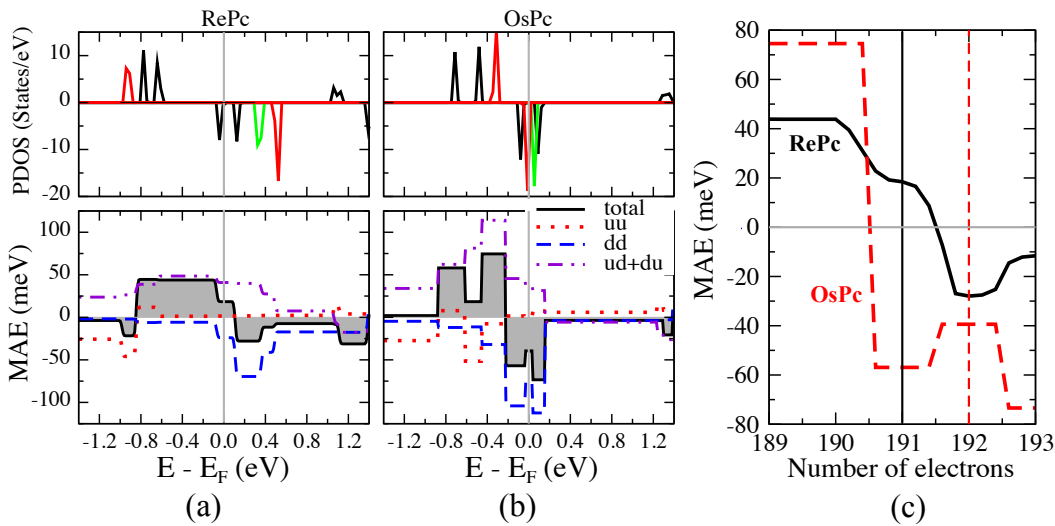

**FIG. 3.** (a) and (b) PDOS of 3d orbitals of Re and Os (upper panels) and total and decomposed MAEs from the torque method with rigid band model (lower panels). Note that the energy levels of $b_{1g}$ orbital are out of the energy range. The zero energy is set to the natural $E_F$ of each TM-Pc molecule. (c) The total MAE as a function of number of valence electrons for RePc and OsPc. The vertical straight lines indicate the number of electrons of neutral TM-Pc molecules.

Nevertheless, the charge state of TM-Pc molecules can be manipulated and the MAE of charged molecules are also of practical interest. To demonstrate the relationship between MAE and electronic structures, we present the projected density of states (PDOS) of different *d*-orbitals of RePc and OsPc molecules in Fig. 3(a) and 3(b), along with the total and decomposed occupancy-dependent MAEs with the rigid band model. It can be seen that the arrangements of the *d*-orbitals are sensitive to their electron occupancies, due to the significant change of the interaction between the TM atom and the surrounding N atoms when the TM atom vary from one to another. As for the $E_F$-dependent MAEs, it can be seen that both the magnitude and sign of MAEs change when the $E_F$ is shifted away for the position for neutral molecule. In particular, giant positive MAE can be obtained for RePc, 44 meV, when the $E_F$ shifts downward by 0.1 eV—corresponding to

[RePc]$^{1+}$ as seen in Fig. 3(c); and for OsPc, 75 meV, when the E$_F$ shifts downward by 0.2 eV—corresponding to and [OsPc]$^{2+}$. In contrast, MAEs of both RePc and OsPc molecules are negative if one shift the Fermi level upwards, i.e., by adding electron(s) to them. Accordingly, the T$_B$ of these positively charged molecules can be much higher than the room temperature (~ 30 meV).

The spin decompositions of MAEs in Fig. 3(a) and 3(b) provide more insights for the manipulation of MAE. For example, the cross spin contributions, MAE(ud+du), are dominant for the neutral RePc molecule, whereas the minority spin contribution, MAE(dd), plays the major role for the neutral OsPc molecule. In this case, MAE(ud+du) remains to be large and MAE(uu) is negligible in a broad energy range, from -0.4 eV to 0.4 eV, so one only needs to reduce MAE(dd) to attain large positive total MAE. This can be done by emptying d-electrons in the minority spin channel, as discussed above for the positively charged RePc and OsPc molecules.

Note that the trend of MAE curves in Fig. 3 can be easily traced to their electronic origins through the following expression [13,33]

$$MAE = \xi^2 \sum_{u,o,\alpha,\beta} (2\delta_{\alpha\beta} - 1) \left[ \frac{|\langle u,\alpha|L_z|o,\beta\rangle|^2}{\varepsilon_{u,\alpha} - \varepsilon_{o,\beta}} - \frac{|\langle u,\alpha|L_x|o,\beta\rangle|^2}{\varepsilon_{u,\alpha} - \varepsilon_{o,\beta}} \right]. \qquad (3)$$

Here $\xi$ is the strength of SOC; $\varepsilon_{u,\alpha}$ and $\varepsilon_{o,\beta}$ are the energy levels of the unoccupied states with spin α ($|u,\alpha\rangle$) and occupied states with spin β ($|o,\beta\rangle$), respectively. L$_x$ and L$_z$ are angular momentum operators along the $x$ and $z$ directions, respectively. Since the e$_g$ orbitals have different magnetic quantum numbers (m=±1) from the $a_{1g}$ (m=0) and $b_{1g}$ (m=±2) orbitals, SOC interaction across these orbitals results in negative MAE(dd), as shown in Fig. 3(a) and 3(b) for the RePc and OsPc molecules. On the other hand, the summations of SOC matrix elements $\langle u,\alpha|L_x|o,\beta\rangle$ and $\langle u,\alpha|L_z|o,\beta\rangle$ also give the orbital moments when the spin aligns along the $x$ and $z$ axes, respectively [34]. This implies that the MAE and orbital moment (M$_L$) usually have the accordant anisotropy [34,35,36]. Indeed, we can see in Fig. 2 that the amplitudes of M$_L$(x) are larger than those of M$_L$(x) for cases with in-plane easy axis, i.e., negative MAEs.

Knowing that [RePc]$^{1+}$ and [OsPc]$^{2+}$ may have large positive MAEs, now the questions is how to achieve the positively charged state for these molecules which actually have appreciable electron affinities and hence prefer negatively charged state [37]. Adsorption of oxidizing atom on the core atom is certainly one of the possible approaches, so we investigated the effect of H, N, O, and F adatoms on top of the RePc and OsPc molecules. Although all atoms except N strongly bind to Re and Os, the strong chemical effect of these adatoms significantly change the arrangements of TM-5d orbitals. As a result, the prediction of MAE from the rigid band model as illustrated in Fig. 3 is not applicable and only H on RePc (H/RePc) gives large positive MAE of 25 meV as listed in table I.

Table I. $M_S$ (in $\mu_B$), $M_L$ (in $\mu_B$), and MAEs (in meV) RePc, RuPc and OsPc with adatom. 'A' stands for adatoms, and 'B' for Re and Os. $M_S$ in parentheses: spin along x and z.

|  | $M_S$ (x) | | | $M_S$ (z) | | | $M_L$ | $M_L$ | MAE |
| --- | --- | --- | --- | --- | --- | --- | --- | --- | --- |
|  | Total | A | B | Total | A | B | (x) | (z) | SCF |
| H/RePc | 1.9 | 0.0 | 1.4 | 1.8 | 0.0 | 1.4 | 0.1 | 0.2 | 25 |
| Os/RuPc | 3.0 | 2.0 | 0.3 | 2.5 | 1.9 | 0.2 | 0.4 | 0.7 | 223 |
| Os/OsPc | 2.2 | 1.6 | 0.1 | 2.6 | 1.9 | 0.1 | 0.4 | 0.7 | 136 |

The alternative way is to use transition metal dimers, as extensively discussed by several authors [35,36,38,39]. Using the same strategy, we place an Os atom on the top of RuPc and OsPc molecules. After structural relaxation, vertical Os-Ru and Os-Os dimers form right in the middle of these molecules, with bond lengths of 2.2 and 2.3 Å, respectively. To further demonstrate the stability of these new Os/RuPc and Os/OsPc molecules, we allow the dissociation of Ru-Os and Os-Os bonds along two pathways, as indicated in Fig. 4. From the energy profiles for Os/RuPc, it can be seen that the removal of Os from the central site of RuPc or OsPc requires high energy costs, 1.7-1.9 eV, so the Os-Re and Os-Os dimers should be extremely stable. Strikingly, the MAEs of Os/RuPc and Os/OsPc are as large as 223 and 136 meV, as listed in table I. These colossal MAEs are sufficient for any technical applications, such as molecular spintronic junction and magnetic storage. To this end, one may need to protect these TM-Pc molecules from interacting with their environment, which is beyond the scope of this work. One encouraging recent experimental progress is that the direct interaction between TM-Pc molecule and substrate can be significantly reduced by placing the molecule on a (2x1) reconstructed Au(110) surfaces [40]. Our predictions should inspire more experimental efforts for the design of innovative magnetic molecules and environments.

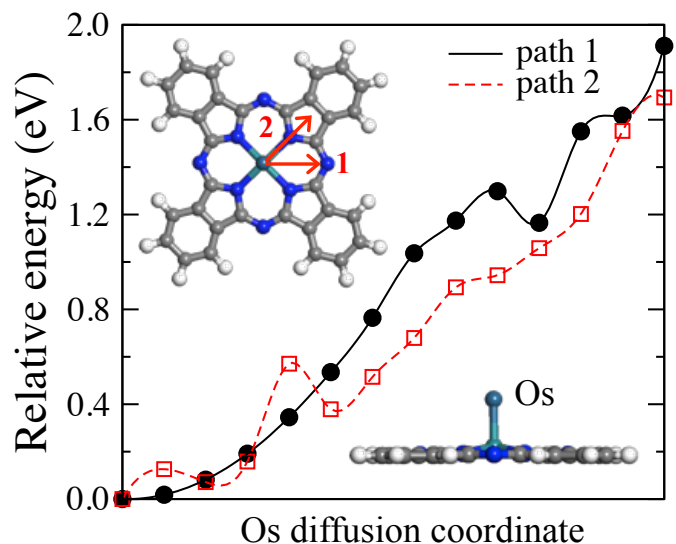

**FIG. 4.** The energy profiles of Os diffusions along the pathways marked in the inset for Os/RuPc. The insets show the top and side views of Os adsorbed on the RuPc molecule. The symbols of atoms are the same as that in Fig. 1.

In summary, we systematically studied the electronic and magnetic properties of the TM-Pc molecules using density functional calculations and found that WPc and RePc have large positive MAEs (~ 22 meV). Based on rigid band model analysis, we elucidated the principles that govern the MAE of TM-Pc molecules and predicted that the positively charged [RePc]$^{1+}$ and [OsPc]$^{2+}$ have huge positive MAEs: 44 and 75 meV, respectively. Strikingly, adsorption of an Os atom above RuPc or OsPc molecule results in stable vertical dimer structures and colossal MAEs: 223 meV for Os/RePc or 136 meV for Os/OsPc. We believe these molecules can serve as the smallest units for magnetic recording and logic operation, and should be very useful in other room temperature molecular spintronics applications.

**Acknowledgements**

Work was supported by DOE-BES (Grant No: DE-FG02-05ER46237) and by NERSC for computing time.